# Tuning of Multicell Superconducting Accelerating Cavities using Pressurized Balloons

Mohamed H. Awida, Donato Passarelli, Marco Castellari, Alessandro Tesi, and Timergali Khabiboulline

*Abstract*—Plastic tuning of multicell superconducting accelerating cavities is crucial in the development cycle of cavities for particle accelerators. Cavities must meet stringent requirements regarding the operating mode frequency, field flatness, and eccentricity before lining them up in a cryomodule string. After dressing bare cavities with helium vessels, the welded vessel prevents access to individual cavity cells disallowing any further localized tuning. Currently, there is no straightforward way to tune dressed cavities other than cutting the vessel and then tuning the bare cavity and dressing it back, which would significantly impact cost and schedule. In this paper, we present a novel tuning technique for already jacketed cavities that is non-invasive and cost-effective. The proposed scheme employs pressurized balloons to be temporarily deployed inside the cavity as a means to localize mechanical deformation in specific cells. The proposed tuning technique was successfully utilized to recover a 9-cell 1.3 GHz tesla-style cavity.

## I. Introduction

SUPERCONDUCITNG radio frequency (SRF) cavities are the core accelerating elements in modern particle accelerator machines. The superconductivity feature allows the cavities to be of extremely low loss exhibiting ultra-high quality factors (~$10^{10}$), which makes them super-efficient in achieving a high voltage gain (~40 MV/m) in a short distance [1]-[2]. Realizing such a high voltage gradient shortens the required length of the particle accelerator, saving hundreds of millions of dollars on infrastructure capital cost. However, in order for each cavity to efficiently accelerate the beam of particles, it has to be tuned in frequency, and the field inside it has to be uniformly distributed in the case of multicell structures [3]-[6]. Cavities are typically made out of multiple cells to improve acceleration efficiency and allow the integration of cavities into cryomodules for particle accelerators [2].

The process of tuning superconducting cavities entails permanently deforming the walls of the cavities by applying appropriate force that exceeds the elastic limit and induces the required plastic deformation [4]. The tuning process typically has three goals; resonance frequency of fundamental mode, field flatness along the various cavity cells of that mode, and alignment of the different cells commonly referred to as eccentricity [5]. The plastic tuning process typically occurs before dressing bare cavities with helium vessels, when the cavity cells are accessible from outside for localized cell tuning. However, unexpected detuning sometimes occurs after dressing during the process of qualifying cavities for string assembly. In principle, a detuned cavity will deem unusable unless tuned back to required specifications.

Previously, the only viable option to recover a detuned jacketed cavity was to cut the helium vessel to have access to the outside walls of the bare cavity, then follow the tuning process of conventional bare cavities. However, cutting the vessel and then welding it back after tuning is an intricate process that incurs significant cost and time.

Conventional tuning of bare superconducting cavities is carried out by applying forces on the outside walls of individual cells, which is obviously not an option for jacketed cavities. However, a similar effect is possible if there is a means to apply localized forces from the inside surface of the cavity. This paper explores the potential of using pressurized balloons to tune jacketed multicell superconducting cavities. The proposed tuning technique is non-invasive and cost-effective.

This paper is organized as follows. In section II, we will briefly discuss the three vital parameters that qualify cavities ultimately for string assembly. Then in section III, we will discuss the conventional tuning techniques before introducing the balloon tuning technique in section IV. Section V will demonstrate the viability of the proposed tuning technique with experimental verification, followed by a conclusion in section VI.

## II. Vitals of a Multicell Accelerating Cavity

Multicell cavities shall meet specific requirements before getting lined up in a cryomodule string to end up operating properly in a particle accelerator machine. Figure 1 depicts, for instance, a 1.3 GHz 9-cell superconducting cavity. Both bare and jacketed cavity versions are shown in (a) and (b), respectively. Three specific requirements are always specified for qualifying cavities to string assembly

Manuscript received Oct 3rd, 2021. This work was supported by the U.S. Department of Energy under Contract Number DE-AC02-07CH11359. Authors are with the Fermi National Accelerator Laboratory, Batavia, IL 60510 USA (630-840-3935; e-mail: mhassan@fnal.gov).

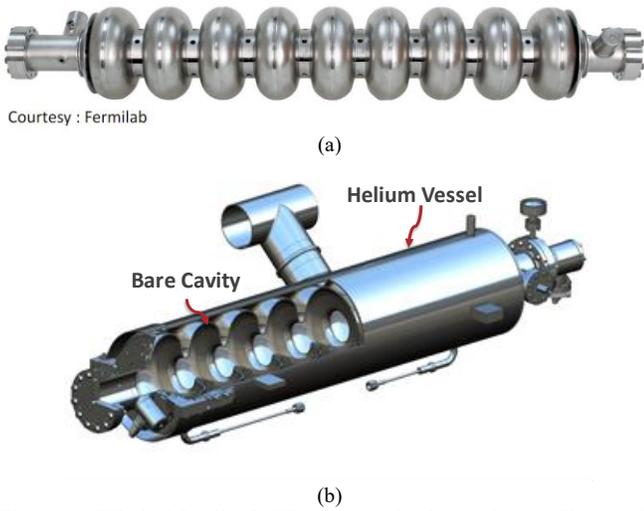

Figure 1. Elliptical 9-cell 1.3 GHz superconducting cavity. (a) Picture of a fabricated bare cavity. (b) Model of a jacketed cavity with a cut away showing the inside of the helium vessel and the niobium bare cavity.

in terms of:

- Resonance Frequency ($f_{pi}$)

Cavities, when at the particle accelerator machine, operate strictly at a particular frequency in a cryogenic environment. Two environmental factors affect the frequencies of cavities between room temperature and the cryogenic environment. One is the process of evacuating the cavities from the air, pumping them down to ultra-high vacuum due to the change in dielectric constant going from air to vacuum. The other factor is the temperature change, which causes thermal shrinkage to the cavity walls and, therefore, a significant frequency shift. Both factors can be accounted for through electromagnetic and mechanical analyses (frequency change due to dielectric constant change can be analytically calculated as frequency is inversely proportional to the square root of dielectric constant).

Moreover, cavities are typically equipped with fast and slow tuners to adjust all cavities to the required frequency of operation in the particle accelerator machine. Nevertheless, the tuners have limited ability to adjust cavities' frequencies (typically a few hundreds of kiloHertz). Therefore, there is an acceptable frequency range for cavities at cryogenic temperature and a corresponding one at room temperature after accounting for the aforementioned environmental factors. It is worth mentioning here also that the type of the tuner and the direction of tuning, whether it is compressing or stretching the cavity, contribute to the acceptable frequency range of cavities. For instance, Fig. 2 demonstrates the performance of an elliptical (TELSA style [7]) 9-cell 1.3 GHz measured at room temperature. The frequency spectrum is shown in Fig. 2(a), demonstrating the nine resonant modes in the first monopole passband, with the operating pi-mode is being the highest in frequency at 1297.8 MHz. We expect about +2 MHz frequency shift during cool down and +0.3 MHz due to cavity evacuation in this particular cavity type. The cavity is projected to land at ~1300.1 MHz when cold. This frequency fulfills the requirements since it is equipped with a compressing tuner that can easily bring the cavity's resonance frequency to the required 1300 MHz frequency of operation.

- Field Flatness (FF)

The second performance vital of a multicell cavity is field flatness (FF) which indicates how the accelerating electric field of the pi-mode is distributed among the various cells along the cavity. Uniform distribution secures better efficiency of particle acceleration. It is typically required that the FF is better than 98% for a bare cavity before dressing and better than 90% for a dressed cavity to get it qualified for string assembly. Figure 2(b) shows the measured field flatness of one of the 1.3 GHz 9-cell cavities during the tuning process. FF is simply the ratio between the minimum to maximum absolute electric field ($E_{min}/E_{max}$ in Fig. 2(b)).

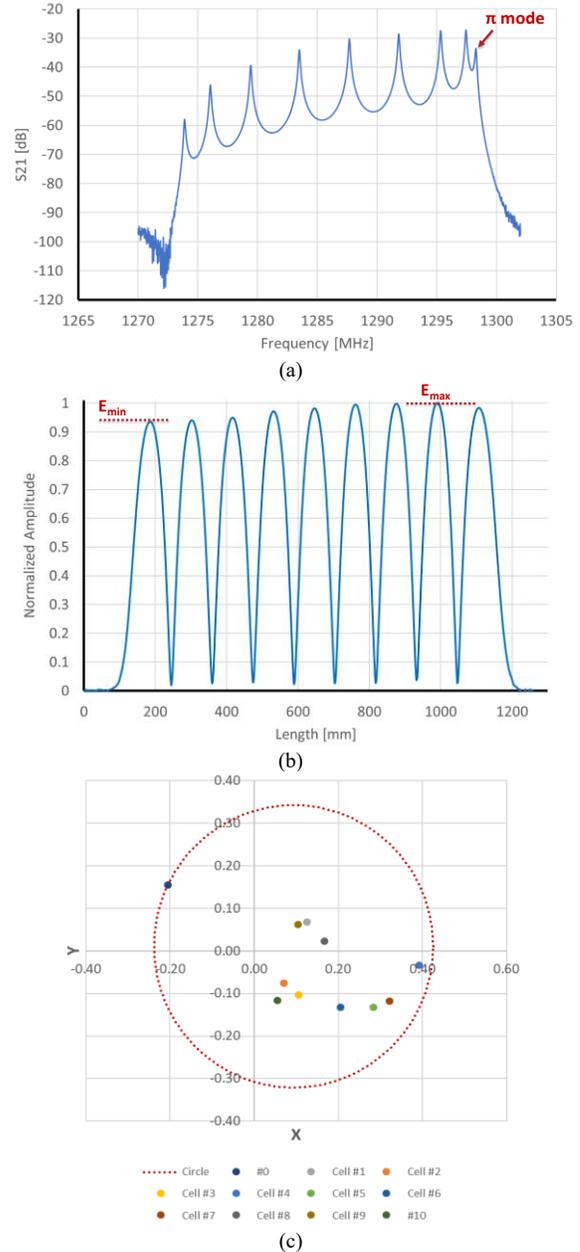

Fig. 2. Performance parameters of an elliptical 9-cell 1.3 GHz superconducting cavity measured at room temperature. (a) RF spectrum demonstrating the nine modes of the first monopole passband with the operating pi-mode is being the highest in frequency. (b) Electric field along cavity axis indicating field flatness ($E_{min}/E_{max}$). (c) Alignment of the various cavity cells with respect to each other showing eccentricity.

- Eccentricity (Ecc)

The third important vital of a multicell cavity is eccentricity (Ecc), which indicates how the various cells are aligned to each other. Eccentricity is measured by basically finding the center of each cell and finding the best fit circle that encloses all the centers of the various cells. An eccentricity of better than 0.5 mm is typically required to qualify cavities for dressing. Figure 2(c) indicates the measured eccentricity of one of the 1.3 GHz cavities after being tuned where the Ecc is better than 0.5 mm.

In that perspective, the process of plastic tuning of superconducting multicell is critical in addressing any issues with these three vitals of a multicell cavity before dressing it to get it qualified for usage in cryomodules, as we will discuss in the next section.

### III. CONVENTIONAL TUNING

Conventional tuning of SRF cavities is typically done by applying localized forces to the individual cells by jaws that close in between the cells at the irises [3]-[6]. Figure 3(a) depicts a cartoon of jaws applied to a cell. Upon stretching (moving the jaws outward) or squeezing (moving the jaws inward) the intended cell beyond the elastic limit, the cell will be plastically deformed, retaining the required deformation when the jaw pressure is released. Stretching a cell entails increasing the frequency of the pi-mode. Moreover, the electric field increases in this particular cell with respect to other cells.

On the contrary, squeezing the cells reduces the electric field and the frequency of the pi-mode. Therefore, resonance frequency and field flatness can be adjusted by stretching or compressing the cells. Meanwhile, eccentricity can be adjusted by applying differential mechanical forces that can move the center of each cell vertically and horizontally.

The tuning process is typically iterative. In the sense that the tuning is done on each cell in the multicell cavity, then the field flatness, frequency, and eccentricity are measured before iterating again on the cells to reach the tuning goals. Meanwhile, several years ago, automatic tuning was devised for 1.3 GHz 9-cell ILC cavities to facilitate the tuning process and make it less labor-intensive [5]-[6]. Figure 3(b) depicts a 1.3 GHz 9-cell ILC cavity on the automatic tuning machine at Fermilab.

### IV. BALLOON TUNING

Cavity tuning, as we explained in the previous section, relies on targeting cells with localized forces applied to the outer wall of the cavity. Once the cavity is jacketed, there is no access to the outer wall. The only possibility to deform a cavity then is to apply force to the cavity's flanges at both ends. We will refer to this force hereafter as "global force". However, applying this global force entails a theoretical uniform distribution of mechanical stresses to all cells with no way to localize these mechanical stresses to a specific cell, which critically limits the tuning capability. In that perspective, the proposed balloon tuning technique strategically addresses this problem. It provides a mechanism to localize the mechanical stresses by using a balloon(s) to be inserted inside the cavity and then pressurized, as shown in Fig. 4. The applied pressure of the balloons provides a local force on the cavity's inner wall that can aid or counter-aid the global force, which results in localized differential force to the targeted cell depending on the balloon location with respect to the other cells that do not have balloons.

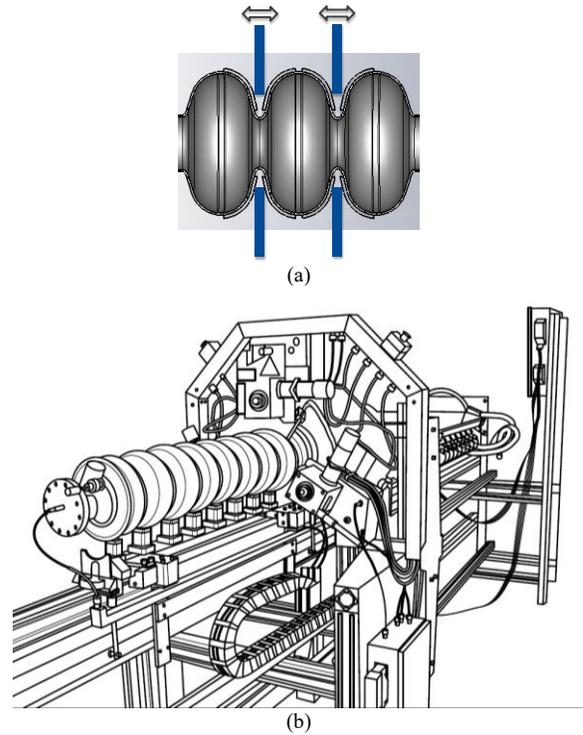

Figure 3. Conventional tuning of a multicell SRF cavity. (a) Cartoon of jaws applied to a cell of the cavity. (b) A 1.3 GHz 9-Cell ILC cavity on Fermilab's automatic tuning machine.

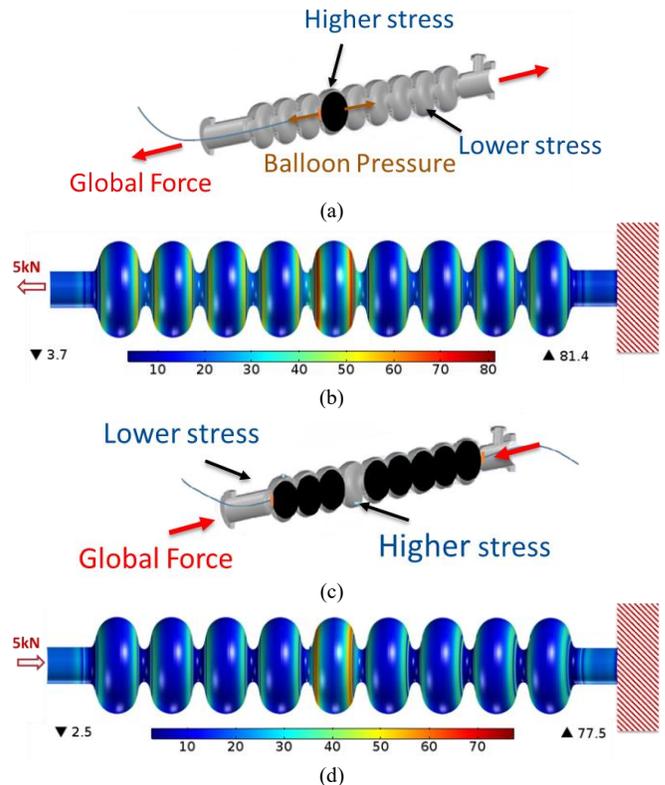

Figure 4. The concept of balloon tuning. (a) Stretching a cell with a pressurized balloon inside the targeted cell in addition to a global stretching force applied to the cavity's flanges. (b) Mechanical analysis of the proposed stretching scheme with 4 bar of pressure in the balloon. (c) Compressing a cell with pressurized balloons on every cell except the targeted cell in addition to the global compressing force applied again to the cavity's flanges. (d) Mechanical analysis of the proposed compression scheme with 4 bar of pressure in the balloons.

Figure 4 demonstrates the concept of balloon tuning in both cases of stretching or compressing a cell in (a) and (c), respectively. The FEM mechanical analyses carried out using Comsol Multiphysics [8] for these two cases are shown in (b) and (d), respectively. In the case of stretching a cell, a single balloon is inserted in the targeted cell then it is pressurized. Upon stretching the whole cavity using a global force on the cavity's flanges, the mechanical stresses in the targeted cell can be increased beyond the yield limit due to the additional local force of the pressurized balloon. Meanwhile, other cells can be managed to stay below the yield limit. The targeted cell thus gets plastically deformed, and the other cells remain in the linear elastic regime. Figure 4(b) shows the results of the mechanical analysis in the case of a global force of 5 kN and a balloon pressure of 4 bar, demonstrating stress of 80 MPa on the targeted cell, while the rest of the cavity is below the 70 MPa yield stress limit of niobium at room temperature. In the case of targeting a cell with compression, we will need to insert balloons on all other cells except the targeted one as shown in (c), then pressurize it and apply a global compression force on the cavity's flanges. The pressurized balloons counter the global compression force preventing all other cells except the targeted one from yielding. Figure 4(d) depicts the mechanical stresses in the case of again a global force of 5 kN and a balloon pressure of 4 bar. It demonstrates how the mechanical stresses are higher in the targeted cell (79 MPa), exceeding the yield limit of niobium at room temperature. Therefore, it secures achieving the plastic deformation of the targeted cell.

## V. EXPERIMENTAL VERIFICATION

We have utilized the proposed balloon tuning technique first to tune a bare 9-cell TESLA style 1.3 GHz cavity, repeating what is possible with the conventional tuning techniques, as a preliminary step, before moving ahead to tune a jacketed cavity. Balloon tuning of the bare cavity went well, fixing its field flatness from 88% to 92.5%. Next, we moved to the more challenging problem of tuning a jacketed cavity. Here we will show the results of this more interesting case. Figure 5 shows the setup we used to tune the cavity in (a), one of the balloons (3-cell) in (b), and two pictures for the cavity's inside from both sides with the balloon inserted in (c) and (d). We picked a jacketed cavity (TB9AES018) that was accidentally deformed during a pressure test. All jacketed cavities are required to pass this pressure test as part of the qualification process. TB9AES018 was significantly deformed during the pressure test because of a lack of proper support. The cavity was deemed unusable for cryomodules as both the frequency and the field flatness were not meeting requirements.

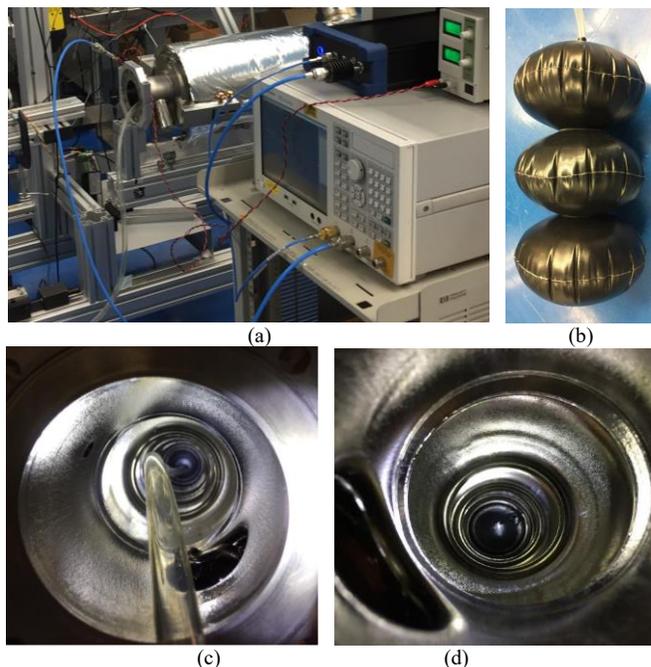

Figure 5. Balloon tuning of a dressed cavity. (a) Setup for jacketed 9-cell ILC cavity (TB9AES018). (b) 3-cell balloons. (c) Picture of the cavity's inside with the balloon inserted (air inlet side) (d) Picture of the cavity from other side.

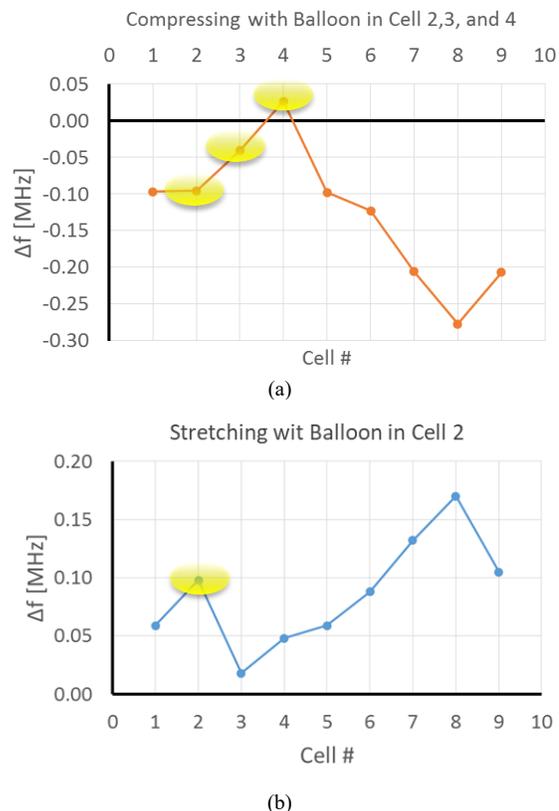

Figure 6. Frequency change per cell during two iterations of balloon tuning of TB9AES018. (a) Compressing the cavity with balloons in cells #2, 3, and 4. (b) Stretching the cavity with a balloon in cell #2.

Figure 6 demonstrates two iterations of the balloon tuning process: one while compressing the cavity with balloons in cells #2, 3, and 4 and a second case while stretching the cavity with a single balloon in cell #2 in (a) and (b), respectively. In the compression iteration, we expect most of the negative frequency change to occur in the cells without balloon support which is demonstrated clearly in Fig. 6(a). Meanwhile, in the case of stretching the cavity, we expect most of the positive frequency change to occur in the cell with the balloon. However, it turned out that cell#8 was quite softer than the other cells, which resulted in a significant shift in this particular cell. Nevertheless, cell #2 also had a positive shift larger than the other cells except for cell #8, which was a good sign that the balloons were working properly to induce the required effect. Despite that having a cell that is quite softer than others complicated the tuning process, as we need to harden it iteratively by stretching then compressing forces, the path to complete the tuning was straightforward. Figure 7 shows the whole process of balloon tuning as both resonance frequency of the pi mode and its field flatness are recorded after each tuning iteration (with an empty cavity after removing the balloons) in (a) and (b), respectively. It shows how we converged on both cavity's vitals, meeting the required specification. The accepted range of both resonance frequency and field flatness is marked by dashed lines in both graphs of Fig. 7. As mentioned before, the softness of cell#8 relatively prolonged the tuning process, but eventually, the cavity was tuned after 24 iterations. Meanwhile, the electric field measured on the cavity axis using the bead pull technique [9]-[10] is shown in Fig. 8 before and after balloon tuning. The cavity, as indicated in Fig. 8, meets LCLS-II [11] project specifications of frequency; 1297.91<f<1298.01MHz and field flatness; FF>90 for it to be qualified for cryomodule string assembly.

On the other hand, after balloon tuning, the cavity underwent a standard chemical preparation process of electropolishing (5 um), pressure rinsing in ultra-pure water, and cleanroom preparation. The cavity was cold tested at 2 K in the vertical test stand at Fermilab. The cryogenic test has shown excellent unloaded quality versus accelerating gradient performance, as shown in Fig. 9. No radiation was observed during the cold test indicating no evidence of field emission. This result verified that the chemical processing successfully removed any residue left on the inner cavity surface from the rubberized nylon balloons.

Meanwhile, the proposed balloon tuning technique was successfully patented [12]-[13]. Future utilization of the proposed tuning technique in other particle accelerator projects is being discussed.

## VI. Conclusion

Utilizing pressurized balloons to apply localized forces to targeted cells of a superconducting multicell jacketed cavity proved viable as a non-invasive tunning technique. Introducing the balloons inside the cavity, then pressurizing it to a 4 bar pressure while applying a global force of about 5 kN on the cavity's flanges provided adequate localized stresses to the targeted cell above the 70 MPa yield limit of niobium at room temperature while the rest of the cavity is managed to stay below the yield limit. Stretching targeted cells is demonstrated by putting the balloons in the targeted cells while applying a global stretching force on the flanges.

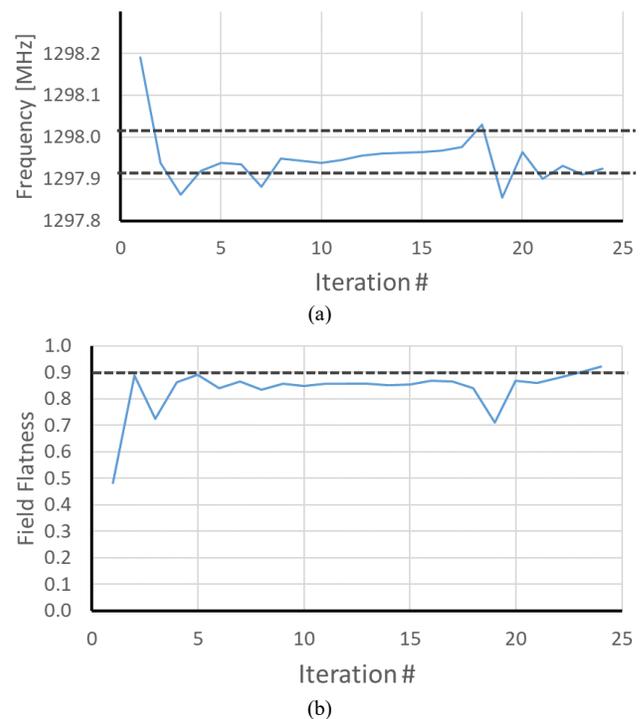

Figure 7. Process of balloon-tuning TB9AES018. (a) Resonance frequency. (b) Field flatness.

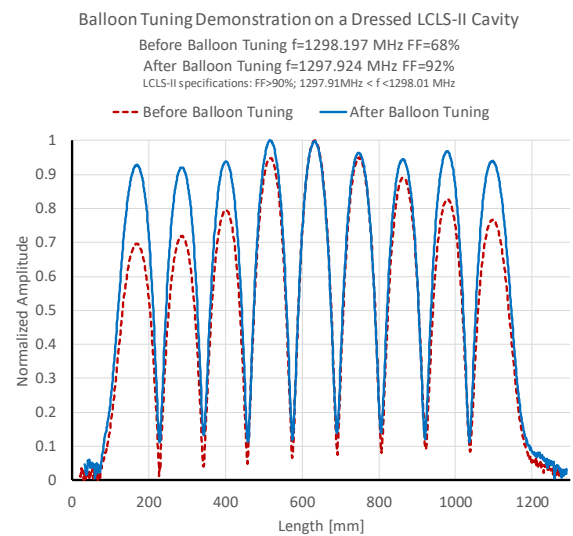

Figure 8. On-axis electric field along TB9AES018 as measured before and after the process of balloon tuning.

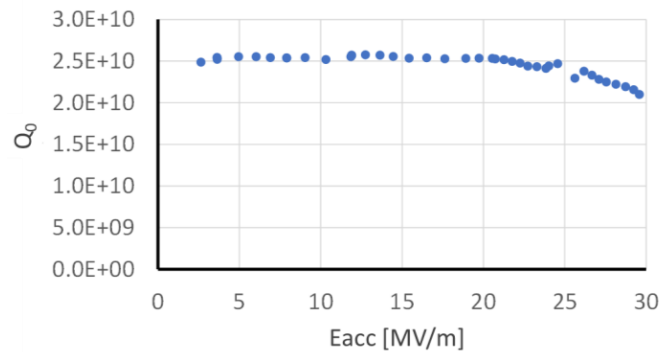

Figure 9. Performance of TB9AES018 in terms of unloaded quality factor versus accelerating gradient at cryogenic temperature (2 K) in the vertical testing dewar at Fermilab.

Similarly, compressing targeted cells is carried out by inserting the balloons in all other cells except the targeted one while applying a global compressing force. Balloon tuning was successfully utilized to tune both bare and jacketed cavities. The jacketed cavity (TB9AES018) was tuned to meet the specifications of cryomodule string assembly in terms of frequency and field flatness. LCLS-II specifications were met after completing the balloon tuning procedure for this cavity. The tuned cavity was successfully tested in the vertical test stand of Fermilab, showing excellent performance in terms of quality factor and absence of radiation. The cold test demonstrated that a typical chemical processing procedure of light electropolishing and pressure rinsing is enough to clean any residues on the cavity's inner surface from the used balloons.

### Acknowledgment

The authors are grateful to the personnel of the Fermilab Argonne joint processing facility and the Fermilab APS-TD SRF Development and Test & Instrumentation departments for their hard work in preparing cavities for cold tests. The authors are also thankful to the vertical testing crew who tested TB9AES018 after balloon tuning.